\documentclass[aps,prb,reprint,showpacs,superscriptaddress]{revtex4-1}

\usepackage{amsmath}
\usepackage{amssymb}
\usepackage{bm,color}
\usepackage{graphicx,epsfig,color}
\begin{document}

\title{Bistable optical response of a nanoparticle heterodimer: 
Mechanism, phase diagram, and switching time}

\author{Bintoro S. Nugroho}
\affiliation{Zernike Institute for Advanced Materials, University of Groningen, Nijenborgh 4, 9747 AG Groningen, The Netherlands}
\affiliation{Jurusan Fisika, Universitas Tanjungpura, Jl. Jendral A. Yani, 78124 Pontianak, Indonesia}

\author{Alexander A. Iskandar}
\affiliation{Physics of Magnetism and Photonics Research Group, Institut Teknologi Bandung, Jl. Ganesa 10, 40132 Bandung, Indonesia} 

\author{Victor A. Malyshev}
\affiliation{Zernike Institute for Advanced Materials, University of Groningen, Nijenborgh 4, 9747 AG Groningen, The Netherlands}

\author{Jasper Knoester}
\affiliation{Zernike Institute for Advanced Materials, University of Groningen, Nijenborgh 4, 9747 AG Groningen, The Netherlands}

\date{\today}

\begin{abstract} 
We conduct a theoretical study of the  bistable optical response of a nanoparticle heterodimer comprised of a closely spaced  semiconductor quantum dot and a metal nanoparticle. The bistable nature of the response results from the interplay between the quantum dot's optical nonlinearity and its self-action (feedback) originating from the presence of the metal nanoparticle. The feedback is governed by a complex valued coupling parameter $G=G_{R}+iG_{I}$. We calculate the bistability phase diagram within the system's parameter space: spanned by $G_\mathrm{R}$, $G_\mathrm{I}$ and $\Delta$, the latter being the detuning  between the driving frequency and the transition frequency of the quantum dot. Additionally, switching times from the lower stable branch to the upper one (and {\it vise versa}) are calculated as a function of the intensity of the driving field. The conditions for bistability to occur can be realized, for example, for a heterodimer comprised of a closely spaced CdSe (or CdSe/ZnSe) quantum dot and a gold nanosphere.
\end{abstract}


\maketitle

\section{Introduction}
\label{Introduction}
Optical bistability is a fascinating nonlinear phenomenon, the essence of which  is  controlling the flow of light by light itself. It is of great importance for optical technologies, in particular, for optical logic and signal processing. The key ingredients for bistable response to occur are optical nonlinearity of the material and a positive feedback. Interplay of the two can result in a multi-valued nonlinear output within a certain range of the system parameter space. A generic optical bistable element exhibits two stationary stable states for the same input intensity, a property which, in principle, opens the door to applications such as all-optical switches, optical transistors, and optical memories.

The phenomenon of optical bistability was predicted by McCall~\cite{mccall1974PRA} in 1974 and demonstrated experimentally for the first time in 1976 by Gibbs, McCall, and Venkatesan~\cite{gibbs1976PRL} (see also Refs.~\onlinecite{lugiantoLA1984PO,gibbs1985B,rosanov1996PO} for an overview). A Fabry-Perot cavity with potassium atoms was used to verify the effect.~\cite{gibbs1976PRL} It has been demonstrated that cavities filled with semiconductor materials as well as semiconductor micro cavities can reveal similar behavior.~\cite{gibbs1979APL,kawaguchi1987OL,gurioli2004PSS,cavigli2005PRB}
 
A vast amount of literature has been devoted to explore the topic (an extensive bibliography can be found in Ref.~\onlinecite{klugkist2007JCP}), especially on the micro- and nanoscale. The development of new (meta-) materials, such as photonic crystals,~\cite{soljacic2003PRO} surface-plasmon polaritonic crystals,~\cite{wurtz2006PRL} and materials with a negative index of refraction,~\cite{litchinitser2007OL} has opened new routes to realize bistable optical elements. Recently, it was suggested that heterodimers of a closely spaced semiconductor quantum dot (SQD) and metal nanoparticle (MNP) would be interesting nanoscale systems that exhibit bistable optical response.~\cite{artuso2008NL,artuso2010PRB,malyshev2011PRB} In fact, such systems have a variety of interesting optical properties that may revolutionarize nanophotonics and optoelectronics.~\cite{brolo2006PCB,viste2010ACN} Amongst these are possible control of the SQD's exciton emission and relaxation properties,~\cite{neogi2004NT,govorov2006NL,neogi2005OL,pons2007NL} nonlinear Fano resonances,~\cite{artuso2008NL,zhang2006PRL,kosionis2012JPC} gain without inversion,~\cite{sadeghi2010NTa} and several other effects.~\cite{sadeghi2009PRB,sadeghi2009NT,sadeghi2010NTb,sadeghi2012APL,hatef2012Nano,
anton2012PRB} All these effects are driven by the strong coupling between  excitons in the SQD and plasmons in the MNP and they are governed by the  geometrical and material parameters of the hybrid cluster, thus providing the perspective to control in detail the optical spectra and dynamics of nanoscale devices.

In this paper, we present an important step in a further understanding of  the optical response of an SQD-MNP heterodimer.  We add bistable to previous work~\cite{artuso2008NL,artuso2010PRB,malyshev2011PRB} a comprehensive analysis of the system's parameter subspace where bistability may occur (the so-called phase diagram), examples of realistic conditions under which bistability may actually be achieved with existing materials (CdSe quantum dot and gold nanoparticle at various distances), a fundamental understanding of the mechanism of bistability, and a study of the switching time of the system between both stable branches. 

With regards to the mechanism of bistability, we focus on the role of the SQD-MNP (complex) coupling parameter $G = G_\mathrm{R} + iG_\mathrm{I}$, which quantifies the self-action (feedback) for the SQD in the presence of the MNP.  We distinguish between the roles of the real and imaginary parts of $G$ ($G_\mathrm{R}$ and $G_\mathrm{I}$) and show that they result in two different mechanisms of the SQD bistability. In the case of $G_\mathrm{R} \not= 0$ and $G_\mathrm{I} = 0$, the feedback is provided by the population-dependent resonance frequency of the SQD, while in the other case, $G_\mathrm{I} \not= 0$ and $G_\mathrm{R} = 0$, it originates from the destructive interference of the driving field with the secondary field produced by the SQD. When $G_{R}\sim G_{I}$, a complicated interplay between both comes into play. We calculate the bistability phase diagram within the system's parameter space spanned by $G_\mathrm{R}$, $G_\mathrm{I}$ and $\Delta$, the latter being the detuning  between the driving frequency and the transition frequency of the quantum dot, and uncover a peculiar behavior of the bistability threshold as a function of $G_\mathrm{R}$ and $G_\mathrm{I}$.  The switching time between both stable branches is calculated as a function of intensity of the driving field, which is important from the viewpoint of practical applications as an all-optical switch. 

This paper is organized as follows. In the next section, we present the system setup and analyze the fields experienced by the SQD and the MNP, both exposed to a driving field. Sec.~III deals with the density matrix formalism for describing the optical dynamics of the  SQD coupled to the MNP.  In Sec.~IV, we discuss in detail the  conditions for bistability to occur in the SQD optical response, based on calculations of the bistability phase diagrams. The physical interpretation of the influence of the SQD-MNP coupling parameter $G = G_\mathrm{R} + iG_\mathrm{I}$ and the detuning away from the SQD resonance is presented.  In Sec.~V, we study the switching time of the system when subjected to a sudden change in the driving intensity. In Sec.~VI, we summarize and conclude.

\section{System setup}
\label{setup}
\begin{figure}[ht]
\begin{center}
\includegraphics[width=0.8\columnwidth]{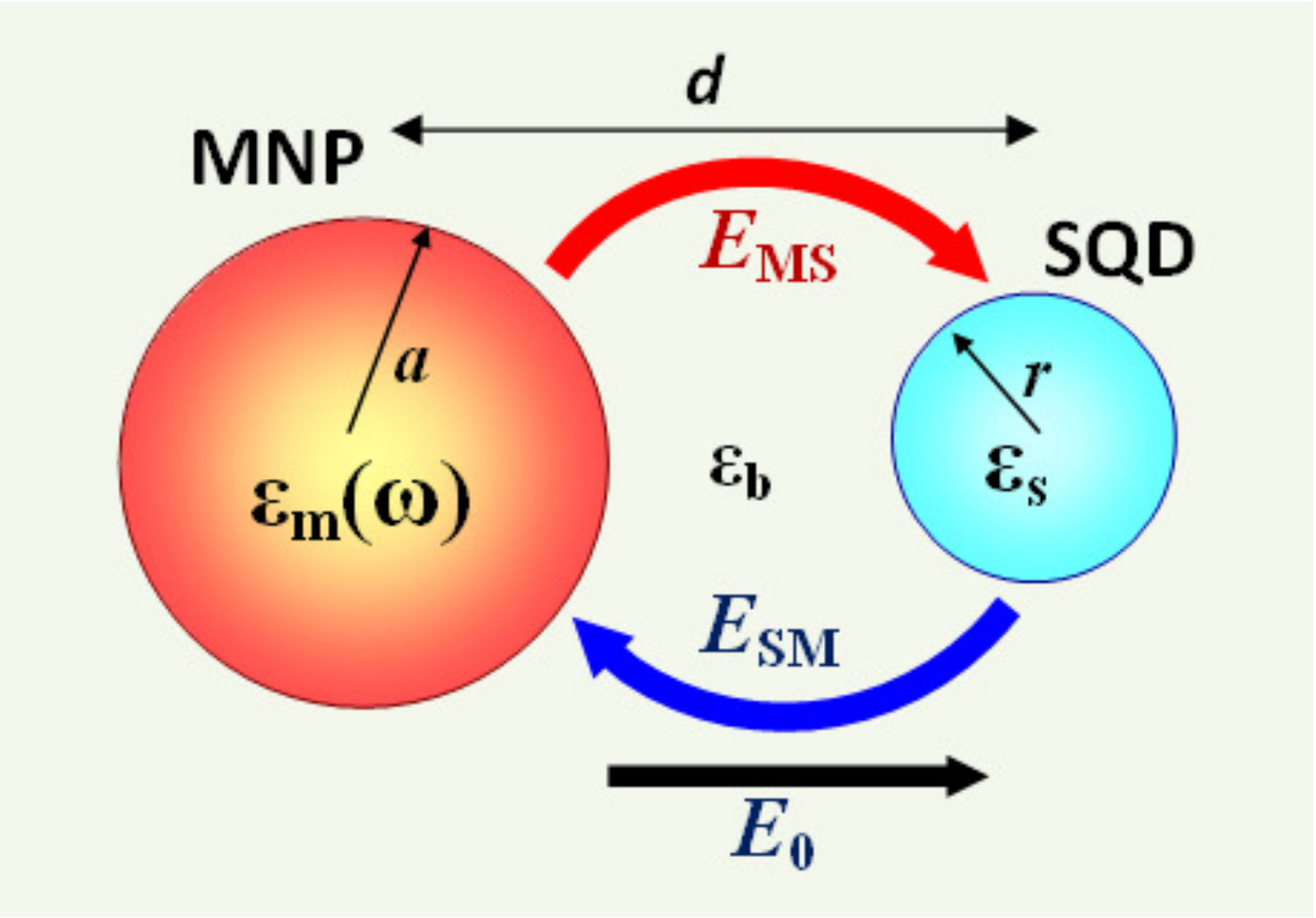}
\end{center}
\caption{Schematics of a SQD-MNP heterodimer embedded in a homogeneous dielectric host with permittivity $\varepsilon_b$ and subjected to an external field of amplitude $E_0$, polarized along the system axis. $\varepsilon_s$ is the SQD bare dielectric constant, $\alpha(\omega)$ is the polarizability of the MNP. $E_\mathrm{SM}$ and $E_\mathrm{MS}$, respectively, denote the electric fields produced by the polarization of the SQD at the position of the MNP and {\it vice versa}.}
\label{Fig1}
\end{figure}

The system of our interest is schematically shown in Fig.~\ref{Fig1}. It is comprised of a single spherical SQD, characterized by a bare dielectric constant $\varepsilon_s$, coupled to a closely positioned spherical MNP with polarizability $\alpha(\omega)$. This heterodimer is assumed to be embedded in a dielectric background with dispersionless permittivity $\varepsilon_b$ and driven by a monochromatic external field $E = (1/2)E_0 \exp(-i\omega\,t) + c.c.$ which is linearly polarized along the SQD-MNP axis. The frequency of the incident field $\omega$ is assumed to be close to the bare exciton transition frequency $\omega_0$ which, in turn, is close to the plasmon resonance peak $\omega_\mathrm{SP}$. We denote the radii of the MNP and the SQD as $a$ and $r$, respectively, while the center-to-center distance between the particles is $d$. These three parameters ($a$, $r$, and $d$) are assumed to be small as compared to the SQD emission wavelength, allowing us to neglect retardation effects and to consider both nanoparticles as point dipoles. 

The dominant optical excitations of the SQD are confined excitons with a discrete energy spectrum. We restrict ourselves to taking into account only one (lowest) exciton energy level characterized by a narrow absorption line width and a transition dipole moment $\mu$. The optical dynamics of the exciton transition will be described quantum mechanically by making use of the Maxwell-Bloch equations for the $2\times2$ density matrix $\rho_{mn}$,  where m and n may be 0 (for the ground state) or 1 (for the excited state). 

The MNP is considered classically in the quasistatic approximation; its response is 
described by the frequency-dependent polarizability $\alpha(\omega)$ within the point dipole approximation (this can be easily generalized to the case of more complex MNP shapes by using an appropriate polarizability tensor). The SQD-MNP interaction will be  treated within the point dipole-dipole approximation.

Now, let us calculate the fields experienced by the SQD and MNP. The external field polarizes the nanoparticles. The polarization of the SQD generates an additional field $E_\mathrm{SM}$  at the position of the MNP and {\it vice versa} $E_\mathrm{MS}$, see Fig.~\ref{Fig1}. These fields are superposed on the external field $E_0$, so that the fields acting upon the SQD and MNP are $E_0 + E_\mathrm{MS}$ and $E_0 + E_\mathrm{SM}$, respectively (acting along the system axes). Note that all above relationships are written for the field amplitudes; the oscillations with the optical frequency $\omega$ already have been extracted. 

Considering the SQD's induced dipole moment $P_\mathrm{SQD}$ as a point dipole, the field $E_\mathrm{SM}$ can be written in the form (see, e.g., Refs.~\onlinecite{Bohren1983B} and~\onlinecite{maier2007B}): 
\begin{equation}
E_\mathrm{SM}=\frac{P_\mathrm{SQD}}{2\pi\varepsilon_{0}\varepsilon_{b}\,d^3} \, .
\label{E_SM}
\end{equation}

Here it is assumed that the SQD is a uniformly polarized sphere, the field of which is screened only by the dielectric constant $\varepsilon_{b}$ of the host medium.~\cite{malyshev2011PRB} Within the density matrix formalism, $P_\mathrm{SQD} = -i\mu R$, where $R$ is the amplitude of the off-diagonal density matrix element $\rho_{10} = -(i/2)R\exp(-i\omega t)$. The MNP dipole moment is now determined by the total field $E_\mathrm{0}+E_\mathrm{SM}$, i.e,   
\begin{subequations}
\begin{equation}
  P_\mathrm{MNP}=\varepsilon_{0}\varepsilon_{b}\alpha(\omega) \left( E_{0} 
  + \frac{P_\mathrm{SQD}}{2\pi\varepsilon_{0}\varepsilon_{b}\,d^{3}} \right) \ , \label{P_MNP}
\end{equation}
\begin{equation}
\alpha(\omega) =4\pi a^{3}\frac{\varepsilon_{m}(\omega)-\varepsilon_{b}}{\varepsilon_{m}
  (\omega)+2\varepsilon_{b}} \ . \label{alpha}
\end{equation}
\end{subequations}
Here, $\varepsilon_m$ is the permittivity of the MNP. The peak of the MNP polarizability $\alpha(\omega)$, when the denominator is minimal, determines the MNP (surface) plasmon resonance. We do not take into account the corrections to $\alpha$ due to the depolarization shift and radiative damping,~\cite{meier1983OL} which are both negligible for the MNP sizes of our interest ($\leq$ 10 nm). The thermal dynamics of the MNP is also neglected: heating of the MNP for the driving field magnitudes of our interest is negligible. 

The field produced by the MNP at the SQD, $E_\mathrm{MS}$, takes the same form as Eq~(\ref{E_SM}), with $P_\mathrm{SQD}$ replaced by $P_\mathrm{MNP}$. The total field experienced by the SQD equals $E_0+E_\mathrm{MS}$. However, the field {\it inside} the SQD should be reduced by an effective SQD dielectric constant $\varepsilon_s^{\prime} = (\varepsilon_s + 2\varepsilon_b)/(3\varepsilon_b)$ (see, e.g., Ref.~\onlinecite{Bohren1983B}, chapter V, page 138, and Ref.~\onlinecite{maier2007B}). Taking all this into account, the total field inside the SQD reads:
\begin{equation}
  E = \frac{1}{\varepsilon_s^{\prime}}\left[1 + \frac{\alpha(\omega)}{2\pi\,d^3} \right]E_0
  + \frac{\alpha(\omega)}{4\pi^2\varepsilon_{0}\epsilon_{b}\varepsilon_s^{\prime}d^{6}}\, 
  P_\mathrm{SQD}.
\label{E}
\end{equation}
Eq.~(\ref{E}) shows two effects for the SQD due to the presence of the MNP. In the first term, one can see a renormalization of the external field amplitude $E_{0}$ by a factor $(1/\varepsilon_s^{\prime})[1+\alpha(\omega)/(4\pi\, d^{3})]$. The second term reveals a self-action of the SQD via the MNP: the field that the SQD experiences, depends on its own state through its dipole moment amplitude $P_\mathrm{SQD}$. As we will show below, this drastically affects the SQD-MNP heterodimer optical response. 

Here, a comment on the second term in Eq.~(\ref{E}) is in order. In a number of recent publications, dealing with the same system, a different formula for this term  was used, in which the factor $\varepsilon_s^{\prime}$ in the denominator appeared squared.~\cite{artuso2008NL,artuso2010PRB,
govorov2006NL,zhang2006PRL,sadeghi2009PRB,sadeghi2009NT,sadeghi2010NTa} We do not agree with this and follow the arguments of Ref.~\onlinecite{malyshev2011PRB} that the above factor should be linear.

\section{Describing the SQD optical dynamics}
\label{SQDdynamics}
As we already mentioned in the previous section, the SQD is assumed to be a two level system, having its filled valence band as ground state $\left|0\right\rangle$ and the lowest exciton level as its excited state $\left|1\right\rangle$; both states are separated by the transition frequency $\omega_{0}$. This approximation is justified when the frequency of the external field $\omega$ is close to the exciton resonance ($\omega \approx \omega_{0}$). Throughout this paper, we use the rotating-wave approximation, so that the time-dependent quantities are the amplitudes of the density matrix elements. The corresponding set of equations reads
\begin{subequations}
\begin{equation}
  \dot{Z} = -\gamma\left(Z + 1\right) - \frac{1}{2}\left(\Omega R^{*}
  +\Omega^{*}R\right) \ , \label{dotZ}\\
  \end{equation}
\begin{equation}
  \dot{R}= -\left(\Gamma +i \Delta\right)R + \Omega Z \ , \label{dotR}
\end{equation}
\end{subequations}
where $Z = \rho_{11}-\rho_{00}$ is the population difference between the excited and ground states of the SQD and $R$ is the amplitude of the off-diagonal density matrix element defined as $\rho_{10}= - (i/2)R \exp(-i\omega t)$. The population difference $Z$ and the amplitude $R$ of $\rho_{10}$ are quantities slowly varying on the scale of an optical period. The constants $\gamma$ and $\Gamma$ represent the rates of population and phase relaxation, respectively, $\Delta = \omega_0 - \omega$ is the detuning away from the SQD resonance, and $\Omega =\mu E_\mathrm{SQD}/\hbar$ is the total electric  field inside the SQD (in frequency units). 

According to Eq.~(\ref{E}), the total field $\Omega$ acting inside the SQD can be written in the form
\begin{equation}
 \Omega = \widetilde\Omega_0 - i\,G\,R \ ,
\label{Omega}
\end{equation}
where $\widetilde\Omega_0$ and $G$ are given by
\begin{subequations}
\begin{equation}
  \widetilde\Omega_0 = \frac{1}{\varepsilon_s^{\prime}} \left[ 1 +
  \frac{\alpha(\omega)}{2\pi\,d^3} \right] \Omega_0 \ , \label{Omega0prime}
\end{equation}
\begin{equation}
  G = \frac{\mu^2\,\alpha(\omega)} {4\pi^2\,\hbar\,\varepsilon_0\,\varepsilon_b\,
  \varepsilon_s^{\prime}\,d^6} \ . \label{G}
\end{equation}
\end{subequations}
Here, $\widetilde\Omega_0$ is the Rabi frequency of the external field renormalized because of the SQD-MNP coupling, with $\Omega_0 = \mu E_0/\hbar$ being the bare Rabi frequency. As we already mentioned in the previous section, the second term in Eq.~(\ref{Omega}) describes the self-action of the SQD via the MNP. The complex-valued constant $G=G_\mathrm{R}+iG_\mathrm{I}$ is a feedback parameter which is determined by the dimer's geometry and material properties. Its real part describes the near-zone feedback field, while the imaginary part is a radiation (far-zone) feedback field (see below). The parameter $G$ contains all information governing the SQD self-action, such as material constants, geometry of the system, and/or details of the interaction ({\it e.g.} contributions of higher multipoles~\cite{yan2008PRB}).

In order to shed light on the effect of self-action on the SQD optical dynamics, we substitute Eq.~(\ref{Omega}) into Eq.~(\ref{dotR}) and obtain
\begin{equation}
  \dot{R} = - \left[\left(\Gamma - G_{I}Z\right) + i\left(\Delta 
  + G_{R}Z\right)\right]R + \tilde{\Omega}_0 Z \ .
\label{dotR1}
\end{equation}
From Eq.~(\ref{dotR1}), it becomes apparent that the SQD self-action has two consequences: (i) - the renormalization of the SQD resonance frequency $\omega_0 \rightarrow \omega_0 + G_\mathrm{R}\,Z$ and (ii) - the renormalization of the dipole dephasing rate $\Gamma\rightarrow\Gamma - G_\mathrm{I}\,Z$; both renormalizations depend on the population difference $Z$. This makes Eqs.~(\ref{dotZ}) and~(\ref{dotR}) nonlinear. Similar renormalizations have been reported in relation with the nonlinear optical response of dense solid state~\cite{hopf1984PRA,ben1986PRA} and gaseous~\cite{friedberg1989PRA} assemblies of two-level atoms, optically dense thin films,~\cite{klugkist2007JCP,basharov1988,benedict1988PRA, benedict1991PRA,malyshev2000JCP} and linear molecular aggregates.~\cite{malyshev1996PRA,malyshev1998PRA} The population dependencies of both the SQD resonance frequency and the dipole dephasing rate provide feedback mechanisms that can give rise to bistability (see below).

\section{Bistability of the optical response}
\label{SQD-MNPbistability}

\subsection{Steady state regime}
\label{SteadyState}
First of all, we are interested in  steady-state solutions of Eqs.~(\ref{dotZ}) and~(\ref{dotR}), which the system reaches after turning on the driving field and waiting until the transient processes are over. Formally, this can be done by setting the time derivatives in Eqs.~(\ref{dotZ}) and~(\ref{dotR}) to zero. After simple algebra, we obtain:
\begin{subequations}
\begin{equation}
    \frac{|\widetilde\Omega_0|^2}{\gamma\,\Gamma}=
    -\frac{Z+1}{Z}\> \frac{|(\Gamma - G_\mathrm{I}Z) + i\,(\Delta 
    + G_\mathrm{R}Z)|^2} {\Gamma^2}, \label{dotZ=0}
\end{equation}
\begin{equation}
    R = \frac{Z\;\widetilde\Omega_0}
	{(\Gamma - G_\mathrm{I}\,Z) + i\,(\Delta + G_\mathrm{R}\,Z)}\ . \label{dotR=0}
\end{equation}
\end{subequations}

As is seen, Eq.~(\ref{dotZ=0}) is a closed equation which is of third order in $Z$. This means that, depending  on the values for $\Delta$, $\gamma$, $\Gamma$, and $G$  it may have three real solutions. The same applies to the dipole moment amplitude $R$. It should be noticed that the possibility of having a three-valued solution to Eq.~(\ref{dotZ=0}) implies three-valued optical response of the SQD-MNP hybrid dimer. However, one branch of the solution turns out to be unstable, as we show below [see Fig.~\ref{Fig2}(b)]. Because of that, we are speaking about bistability (not tristability).

\subsection{General study: Bistability phase diagram}
\label{BistabPhaseMap}

It is of interest to perform a general study of the system's bistability, examining the occurrence of the effect in the parameter space $G_\mathrm{R}$, $G_\mathrm{I}$, $\Delta$, and $\Gamma$. As follows from Eq.~(\ref{dotZ=0}), the relaxation constant $\Gamma$ can be used as a unit for $G_\mathrm{R}$, $G_\mathrm{I}$, and $\Delta$ and thus is not a relevant parameter. 

Our study is based on Eq.~(\ref{dotZ=0}) which is of the third order in $Z$. Therefore, this equation may have three real roots for specific values of $G_\mathrm{R}$, $G_\mathrm{I}$, and $\Delta$. The solutions are different when $|\widetilde\Omega_0|^2/(\gamma\,\Gamma)$ in Eq.~(\ref{dotZ=0}), formally considered as a function of $Z$, has a minimum and maximum. The threshold for bistability is determined by the condition that the derivative of $|\widetilde\Omega_0|^2/(\gamma\,\Gamma)$ with respect to $Z$ has a degenerate root (merged extrema). We used this definition to calculate the bistability phase diagram. 

\begin{figure}[ht]
\begin{center}
\includegraphics[width=0.47\columnwidth]{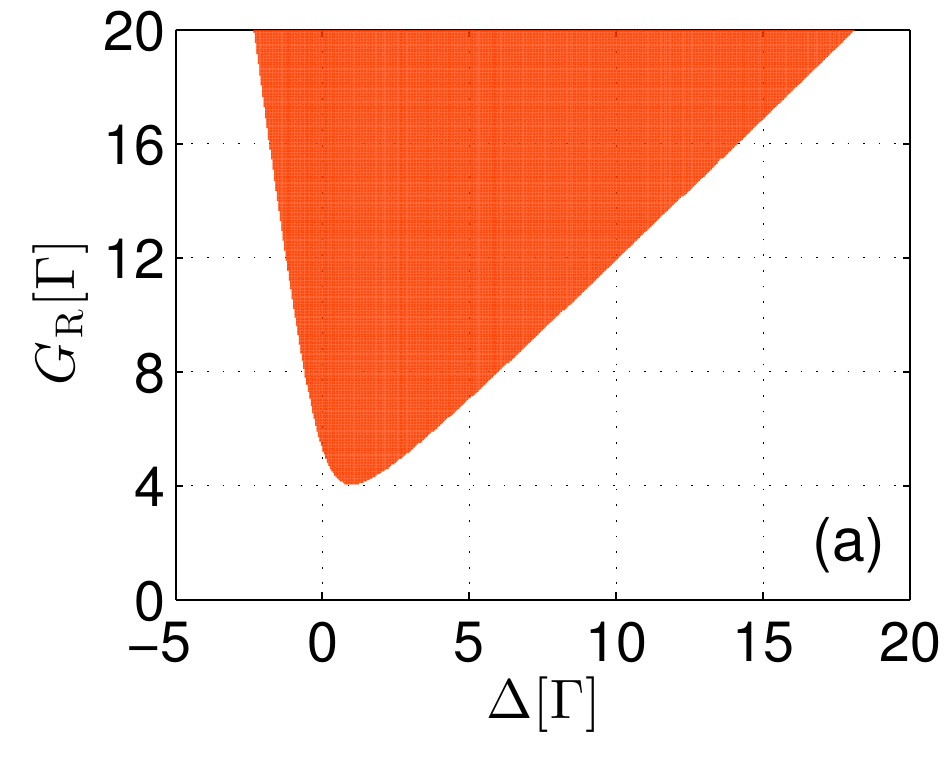}
\includegraphics[width=0.47\columnwidth]{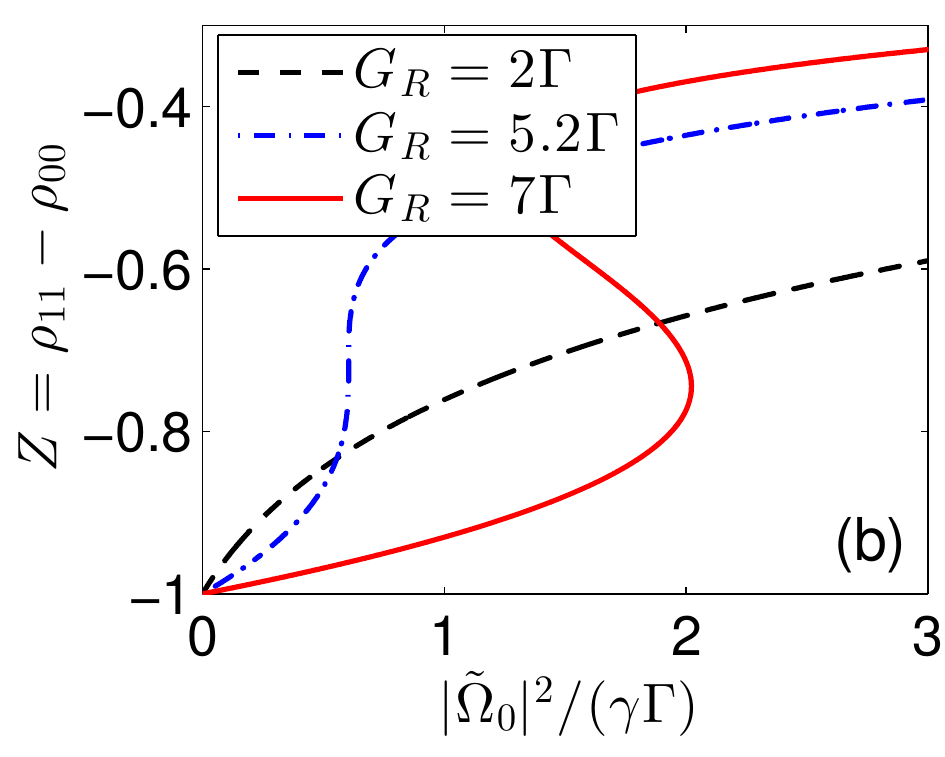}
\includegraphics[width=0.80\columnwidth]{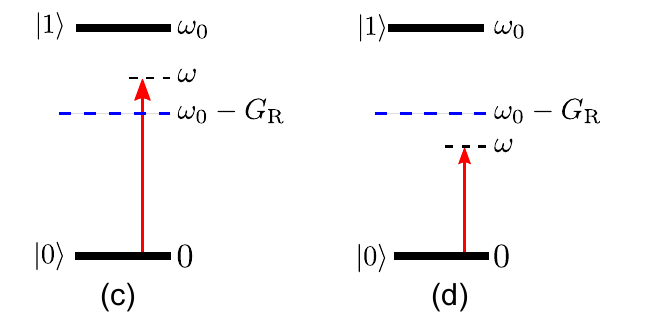}
\end{center}
  \caption{(a)~Bistability phase diagram of the SQD optical response in 
  the parameter subspace $[G_\mathrm{R};\Delta;G_\mathrm{I}=0]$. The colored area
  shows the parameter subspace where bistability exists. The boundaries between white and colored regions represent the bistability threshold within the corresponding parameter sub-space. (b)~Steady state solution of Eq.~(\ref{dotZ=0}) with $\Delta = 3\Gamma$ for various $G_\mathrm{R}$ (see the legend). (c)~Schematics of excitation when bistability may occur ($\omega > \omega_{0}-G_{R} $) and (d)~when bistability does not occur at all ($\omega < \omega_{0}-G_{R} $).}
\label{Fig2}
\end{figure}

\begin{figure}[ht]
\begin{center}
\includegraphics[width=0.47\columnwidth]{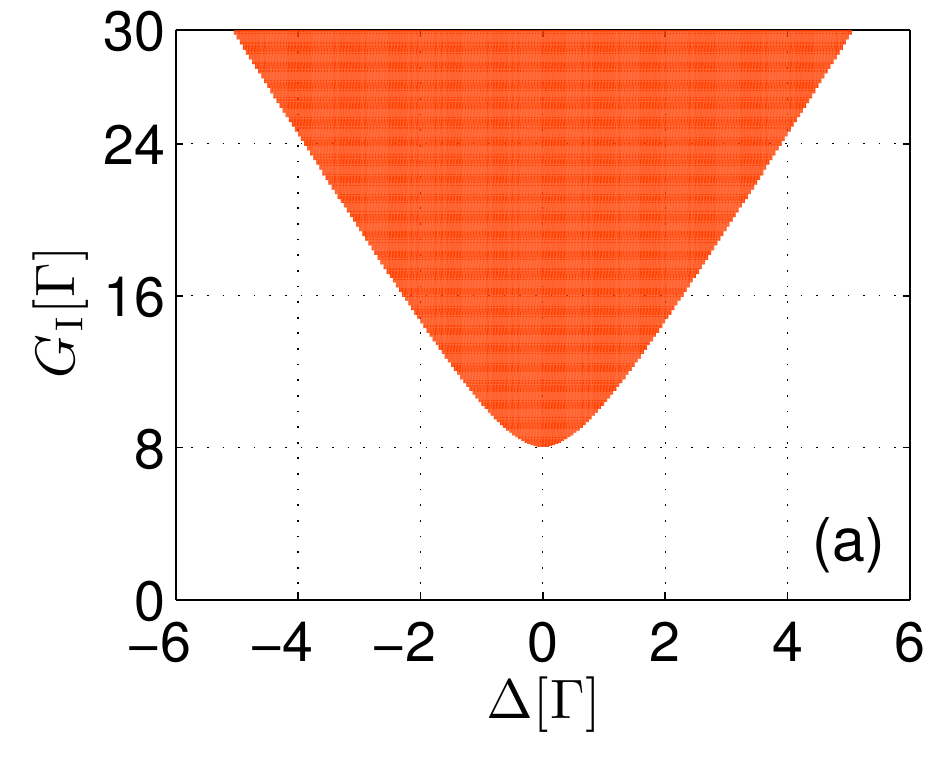}
\includegraphics[width=0.47\columnwidth]{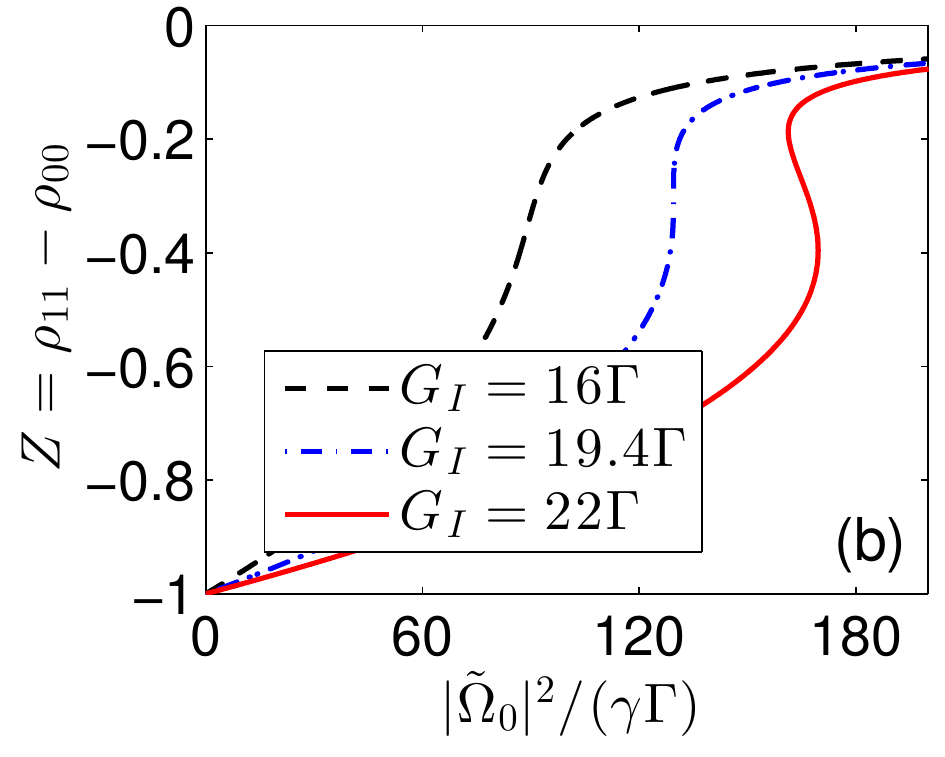}
\includegraphics[width=0.80\columnwidth]{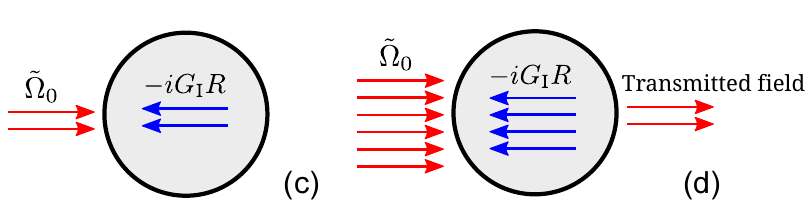}
\end{center}
  \caption{(a)~Bistability phase diagrams of the SQD optical response in 
  the parameter subspace $[G_\mathrm{I};\Delta;G_\mathrm{R}=0]$. The colored area
  shows the parameter subspace where bistability exists. The boundaries between white and colored regions represent the bistability threshold within the corresponding parameter sub-space. (b)~Steady state solution of Eq.~(\ref{dotZ=0}) for $\Delta = 3\Gamma$ and several $G_\mathrm{R}$ (see the legend). (c)~Schematics of the total field in the SQD at low excitation when $-iG_\mathrm{I}R$ compensates ${\widetilde \Omega}_0$ and (d) when the SQD is saturated.}
\label{Fig3}
\end{figure}

To understand the role of the real and the imaginary part of the feedback parameter $G$, Eq.~\ref{G}, in the mechanism of bistability, we calculate the phase diagram setting $G_I =0$, whereas $G_\mathrm{R} \ne 0$ and $\Delta \ne 0$, and $G_\mathrm{R} = 0$, but now $G_\mathrm{I} \ne 0$ and $\Delta \ne 0$, respectively. The results are shown in Fig.~\ref{Fig2} and Fig.~\ref{Fig3}. Colored areas denote the parameter sub-space where bistability occurs and the boundaries between white and colored regions represent the bistability threshold for given parameters.   

Figure~\ref{Fig2}(a) shows the bistability phase diagram within the parameter sub-space $[G_\mathrm{R};\Delta;G_\mathrm{I} = 0]$. First, we observe that there is an absolute threshold for the occurrence of bistability  with respect to $G_\mathrm{R}$: the effect exists only if $G_\mathrm{R} > 4 \Gamma$. This is in agreement with the analytical result derived by Friedberg et al.~\cite{friedberg1989PRA} for a dense gaseous medium. 

In Fig.~\ref{Fig2}(b) we also present the solutions of Eq.~(\ref{dotZ=0}) with $\Delta = 3\Gamma$ for $G_\mathrm{R}$ below, above, and at exactly the bistability threshold. As is seen, for $G_\mathrm{R} = 2\Gamma$, (below the bistability threshold) the dependence of the SQD population difference $Z$ on the external field intensity, $|\tilde{\Omega}_0|^2/(\gamma\Gamma)$, is single valued (bistability does not occur). At $G_\mathrm{R} = 5.2\Gamma$, one observes an inflection in the $Z$-vs-intensity dependence, which denotes that the derivative of $Z$ with respect to $|\tilde{\Omega}|^2/\gamma\Gamma$ has degenerate root. For the higher value of $G_\mathrm{R} = 7\Gamma$, Eq.~(\ref{dotZ=0}) reveals a three valued solution, and the population-vs-intensity curve has an S-like shape: a signature of the bistable behavior.     

The bistability mechanism in the present case (when $G_\mathrm{I} = 0$ or $G_\mathrm{R} \gg G_\mathrm{I}$) is similar to the one known for a thin film of two-level atoms, where the feedback is provided by the Lorentz-Lorentz local field. In the case of a SQD-MNP nanodimer, the field produced by the MNP plays a role of a local field. As is seen from Eq.~(\ref{dotR1}), the feedback, originating from $G$, gives rise to a population dependence of the SQD resonance frequency via the $G_\mathrm{R}Z$ term; the resonance will be red-shifted (renormalized) to $\omega_0 + G_\mathrm{R}Z$, ranging from $\omega_0 - G$ to $\omega_0$ (remember that under steady state conditions, $-1 \ge Z \le 0$, i.e., $Z$ is negative, whereas we assume that $G_\mathrm{R} > 0$). As the population difference $Z$ grows (become less negative) when increasing the applied intensity $|\tilde{\Omega}_0|^2/(\gamma\Gamma)$, the renormalized resonance frequency $\omega_0+G_\mathrm{R}Z$ approaches $\omega_0$. Thus when the detuning $\Delta$ falls within the window
$[\omega_0 -G_\mathrm{R},\omega_0]$ ($\Delta<G_\mathrm{R}$), the incident intensity will bring the system closer to resonance. This underlies the occurrence of bistability [see Fig.~\ref{Fig2}(c)]. Increasing the detuning $\Delta$ requires a larger $G_\mathrm{R}$ to get bistable response. When $\Delta$ is outside the window $[\omega_0 -G_\mathrm{R},\omega_0]$ ($\Delta > G_\mathrm{R}$), the excitation drives the system out off resonance upon increasing the incident intensity [see Fig.~\ref{Fig2}(d)]. 

Note that the phase diagram in the present case is strongly asymmetric with respect to changing $\Delta$ to $-\Delta$. The reason is that at a positive detuning ($\omega_0 > \omega$), the SQD can get in resonance with the external field: tuning the population difference within $-1 < Z < 0$ allows this. At a large negative detuning ($\omega_0 < \omega)$), the situation is different: the resonance condition requires a significant positive population difference $Z$, which is unreachable under steady state excitation. 

Within the parameter sub-space $[G_\mathrm{I};\Delta;G_\mathrm{R} = 0]$ the absolute threshold for bistability turns out to be $G_\mathrm{I} = 8\Gamma$ [see Fig.~\ref{Fig3}(a)]. At smaller $G_\mathrm{I}$, the effect is absent. This result can be verified analytically by setting $G_\mathrm{R} = \Delta = 0$ in~(\ref{dotZ=0}) and analyzing the derivative of $|\widetilde\Omega_0|^2/(\gamma\,\Gamma)$ with respect to $Z$, as explained above.~\cite{malyshev2000JCP} Increasing the detuning $\Delta$ requires larger values of $G_\mathrm{I}$. However, unlike the previous sub-set of parameters $[G_\mathrm{R};\Delta;G_\mathrm{I}=0]$, the bistability phase diagram here is symmetric upon changing the sign of $\Delta$, as is also evident from Eq.~(\ref{dotZ=0}). In Fig.~\ref{Fig3}(b), plots of the solutions to Eq.~(\ref{dotZ=0}) are presented for $G_\mathrm{I} = 16\Gamma$ (below the threshold), for $G_\mathrm{I} = 22\Gamma$) (above the threshould), and for $G_\mathrm{I} = 19.4\Gamma$ (exactly at the threshold). 

The mechanism of bistability when only $G_\mathrm{I}$ plays a role ($G_\mathrm{R} \ll G_\mathrm{I}$) is as follows. As is mentioned in Sec.~\ref{SQDdynamics}, the total field acting inside the SQD is given by Eq.~(\ref{Omega}). Using Eq.~(\ref{dotR=0}), it can be easily shown, that at a low level of excitation ($Z \approx -1$), the feedback field $-iG_\mathrm{I}R$ is out of phase with the external one $\widetilde{\Omega}_0$, and at $G_\mathrm{I} \gg \Gamma$ almost compensates the latter [see Fig.~\ref{Fig3}(c)]. The total field $\Omega = \widetilde{\Omega}_0 - iG_\mathrm{I}R$ is on the order of $\Gamma/G_\mathrm{I}$~\cite{klugkist2007JCP,malyshev2000JCP}, i.e., is very small, thus preventing bistability to occur. As the system is being excited, the compensation decreases, leading to an increase of the total field $\Omega$ inside the SQD and saturating the SQD transition [Fig.~\ref{Fig3}(d)]. This is the why, in this case, bistability occurs at higher intensity [compare Fig.~\ref{Fig2}(b) and Fig.~\ref{Fig3}(b)]. Finally, such an interference-based mechanism also gives rise to the second self-sustaining stable state (the upper branch), provided $G_\mathrm{I}$ is sufficiently large.

\begin{figure}[ht]
\begin{center}
\includegraphics[width=0.49\columnwidth]{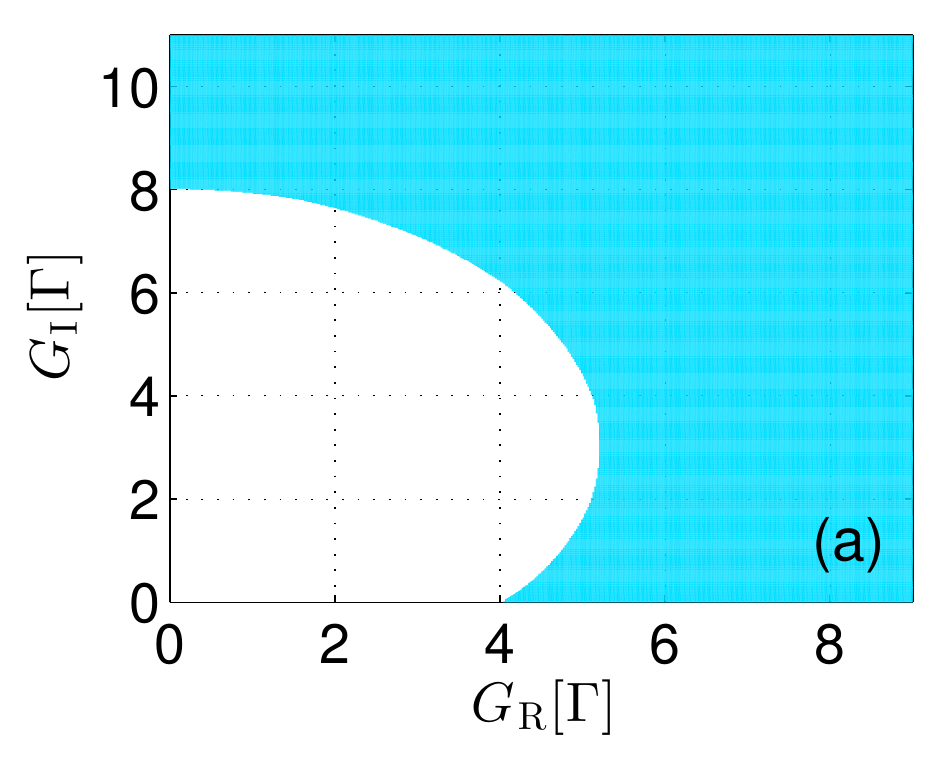} \includegraphics[width=0.49\columnwidth]{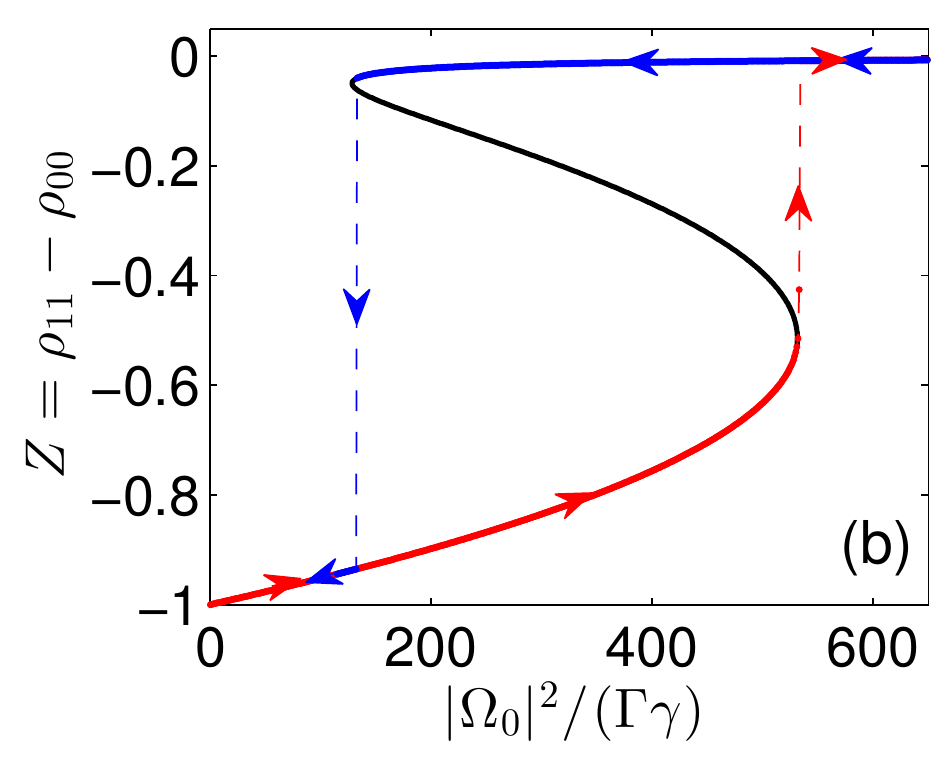}
\end{center}
\caption{(a)~Bistability phase diagram of the SQD optical response in the parameter subspace $[G_\mathrm{R};G_\mathrm{I}]$. The colored area shows the parameter subspace where bistability exists. The boundaries between white and colored regions represent the bistability threshold within the corresponding parameter sub-space. (b)~Steady state solution of Eq.~(\ref{dotZ=0}) (black line) and the solution of Eqs.~(\ref{dotZ}) and (\ref{dotR}) under adiabatic sweeping up and down of the external field intensity $|\Omega_0|^2/(\gamma\Gamma)$.
The arrows show the system’s time domain route (hysteresis loop), indicating that only two branches of the S-shaped $Z$-vs-intensity characteristics are stable, whereas the intermediate branch (black curve) in unreachable (unstable).  The set of parameters is described in the text.
}
\label{Fig4}
\end{figure}

Fig.~\ref{Fig4}(a) shows the bistability phase diagram within the parameter sub-space [$G_\mathrm{R};G_\mathrm{I};\mathrm{any}\,\Delta$] (see caption for explanation). The diagram is symmetric with respect to the transformation $G_\mathrm{R}$ to $-G_\mathrm{R}$. Because of that, we present it only for $G_\mathrm{R} > 0$.~\cite{endnote}  Remember that the colored area shows the range of $G_\mathrm{R}$ and $G_\mathrm{I}$ where bistability may exist. Considering the diagram, we make several observations. First, the absolute bistability threshold at $G_\mathrm{I} = 0$ is $G_\mathrm{R} = 4\Gamma$, whereas at $G_\mathrm{R} = 0$, it is $G_\mathrm{I} = 8\Gamma$, in accordance with the results presented in Figs.~\ref{Fig2}(a) and~\ref{Fig3}(a). Second, within the range $0 < G_\mathrm{R} < 4\Gamma$ (below the bistability threshold with respect to $G_\mathrm{R} = 4\Gamma$ at $G_\mathrm{I} = 0$), the bistability threshold with respect to $G_\mathrm{I}$ decreases from $G_\mathrm{I} = 8\Gamma$ to $G_\mathrm{I} = 6.2\Gamma$, meaning that increasing $G_\mathrm{R}$ within this range promotes the occurrence of bistability. Finally, when $4\Gamma < G_{R} < 5.2 \Gamma$ (above the bistability threshold with respect to $G_\mathrm{R} = 4\Gamma$ at $G_\mathrm{I} = 0$), there is a range of $G_\mathrm{I}$ values, depending on the value of $G_\mathrm{R}$, where bistability does not exist. The presence of this area in the phase diagram originates from the complicated interplay of two fields: the renormalized external field $\widetilde{\Omega}_0$ and the feedback field $-iG\,R$, which both determine the total field $\Omega$ inside the SQD, see Eq.~(\ref{Omega}). Within this area, these two fields interfere destructively with each other, thus preventing the occurrence of bistability. 

As an example, consider a system consisting of a CdSe SQD coupled to a gold MNP. In our numerical calculations, the following set of the SQD parameters was selected: the transition energy $\hbar\omega_0 = 2.36$ eV (which corresponds to the optical transition in a 3.3 nm radius SQD), the transition dipole moment $\mu = 0.65$ $e\cdot$ nm, the SQD bare dielectric constant $\varepsilon_{s} = 6.2$, the host dielectric permittivity $\varepsilon_{b} = 1$, and the SQD relaxation constants $\gamma = 1.25$ ns$^{-1}$ and $\Gamma = 3.33$ ns$^{-1}$.~\cite{zhang2006PRL} We chose the MNP radius $a = 10$ nm, the MNP-SQD center-to-center distance $d = 17$ nm, and the bare exciton detuning $\Delta=\Gamma$. The tabulated data for the permittivity of gold $\varepsilon_m(\omega)$~\cite{johnson1972PRB} have been used to calculate the MNP polarizability $\alpha(\omega)$, according to Eq.~(\ref{alpha}). We found that $\alpha(\omega)$ has a peak at $\hbar\omega_\mathrm{SP} = 2.4$ eV with a width on the order of 0.25 eV (see also Ref.~\onlinecite{zhang2006PRL}). These data allowed us to extract the feedback parameter $G$ using Eq.~(\ref{G}). As is seen from this equation, $G$ is a function of frequency. However, the frequency domain of our interest is determined by a narrow region around the SQD sharp resonance (at most of the order of 10 $\Gamma$, see below), whereas the MNP plasmon peak is much broader. Therefore, $G$ is required just at the SQD resonance frequency $\omega_0$. At this frequency, for the set of parameters used, $G = (27.1 + 11.1i)\Gamma$ is well inside the bistability region [see Fig.~\ref{Fig4}(a)]. 

Using the above set of parameters and choosing the bare detuning away from the SQD resonance $\Delta = \Gamma$, we solved Eqs.~(\ref{dotZ}) and (\ref{dotR}) numerically under adiabatic sweeping up and down of the external field intensity $|\Omega_0|^2/(\gamma\Gamma)$ and obtained a hysteresis loop of the SQD optical response, presented in Fig.~\ref{Fig4}(b). The arrows show the system´s time domain route, indicating that the intermediate branch (black curve with negative gradient) is unreacheble (unstable) when adiabatically sweeping the incoming field intensity.

\section{Switching Time}
\label{switching}

\begin{figure}[ht]
\begin{center}
\includegraphics[width=0.8\columnwidth]{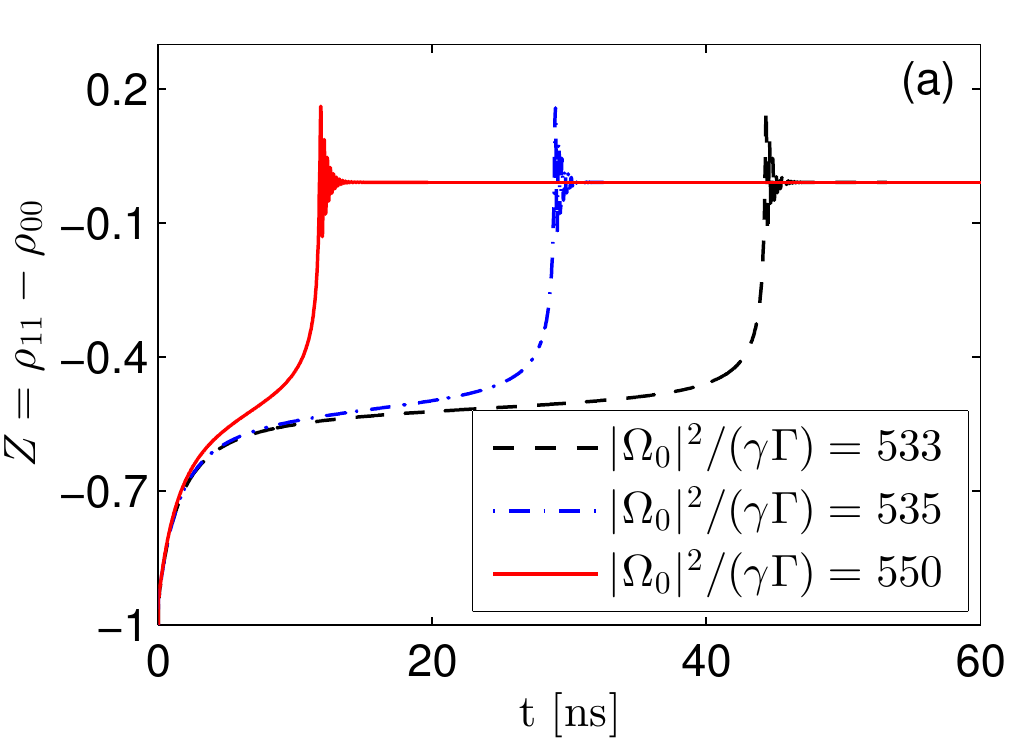}
\includegraphics[width=0.8\columnwidth]{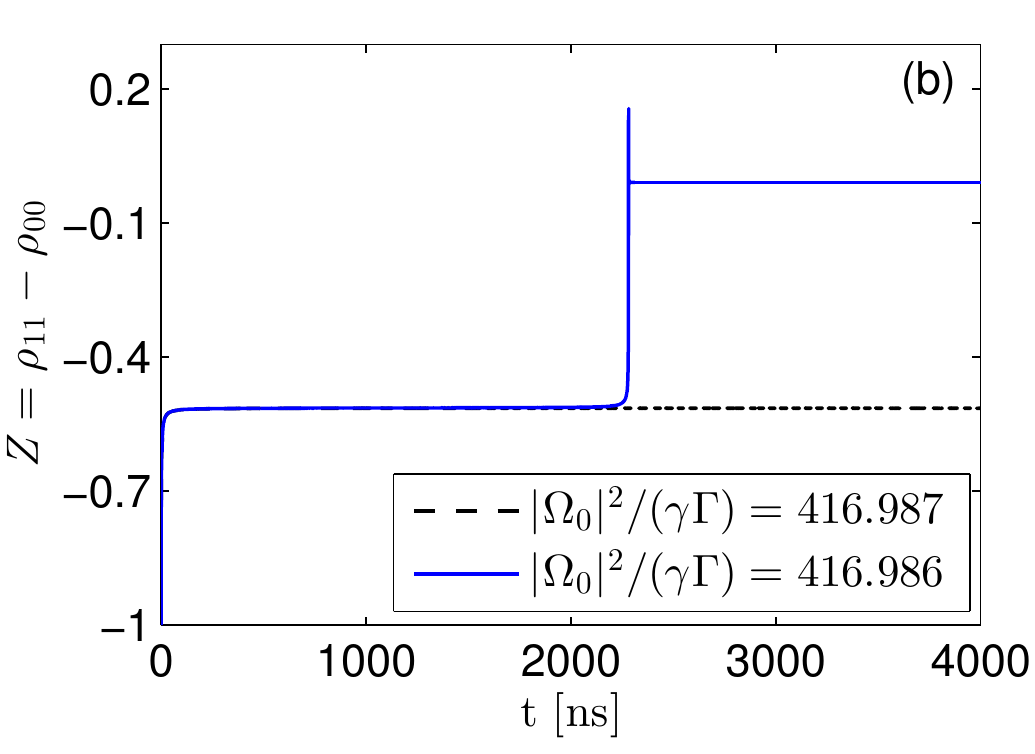}
\end{center}
\caption{(a)~Time evolution of the SQD population difference $Z$ after suddenly switching the incident intensity $I_{0}$ from zero to a value above the upper critical one $I_\mathrm{c} = |\Omega_\mathrm{0c}|^2/(\gamma\Gamma) = 531.4125$ of the bistable $Z$-vs-$I_\mathrm{0}$ characteristics in Fig.~\ref{Fig4}(b). (b)~As in (a) but now $I_0$ is in close proximity to $I_c$, demonstrating the slowing down of the population relaxation when approaching the critical point from the above [compare the time scales in (a) and (b)]. The system parameters were taken as in Fig.~\ref{Fig4} (b).}
\label{Fig5}
\end{figure}

\begin{figure}[ht]
\begin{center}
\includegraphics[width=0.8\columnwidth]{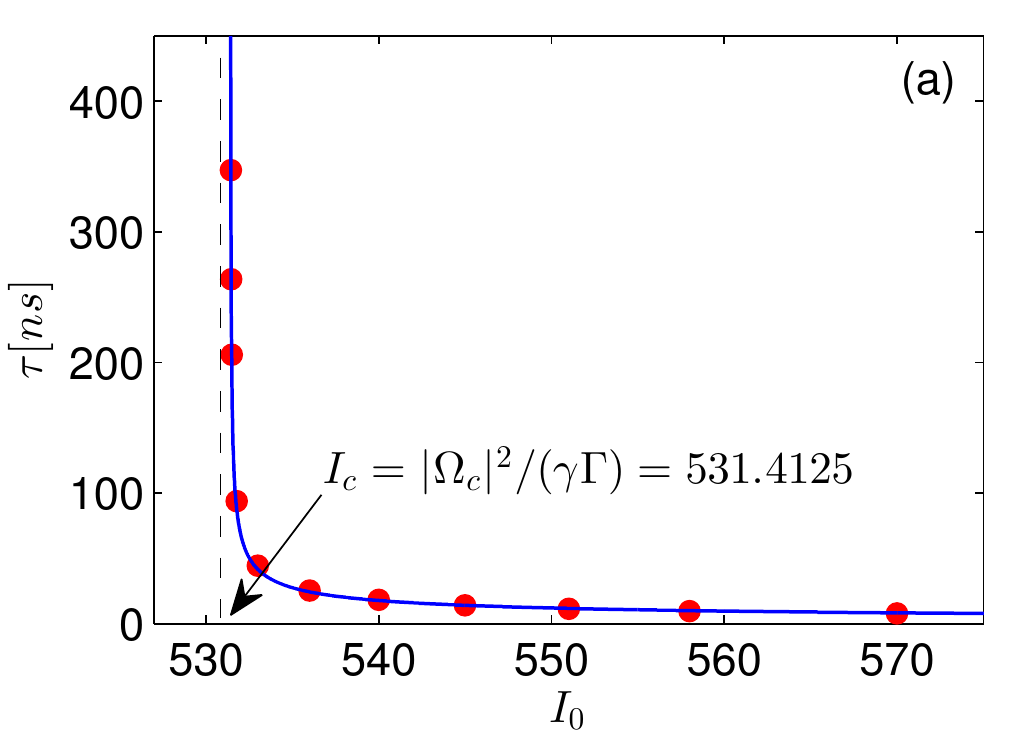}
\includegraphics[width=0.8\columnwidth]{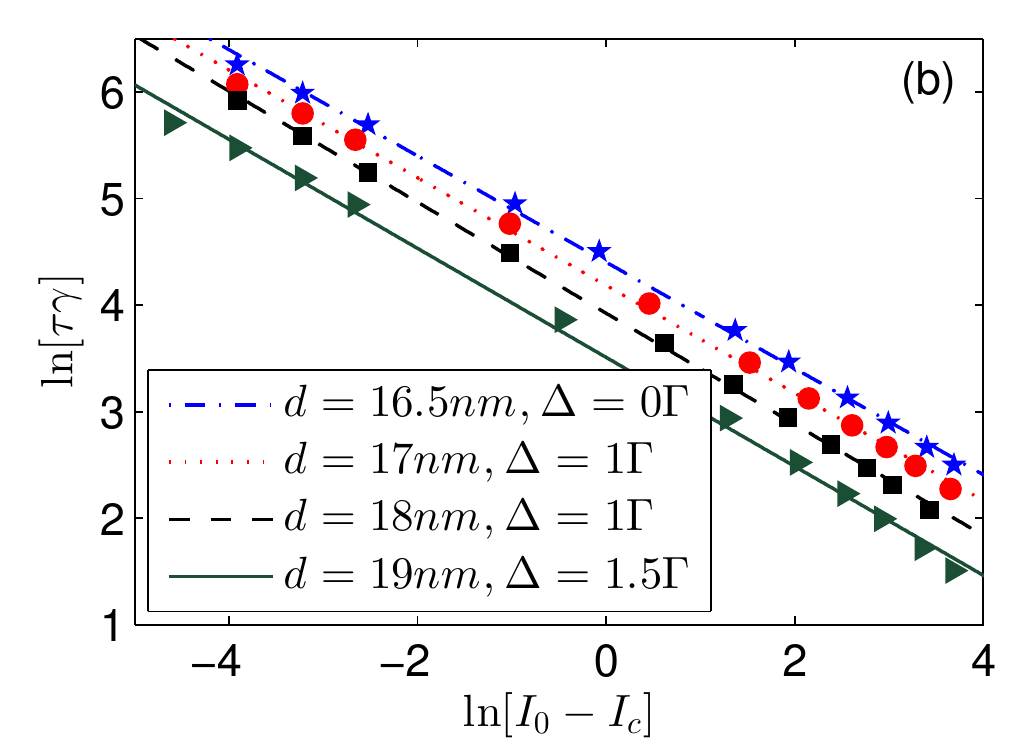}
\end{center}
\caption{(a) Relaxation time $\tau$ of the population difference $Z$ as a function of the incident intensity $I_\mathrm{0} = |\Omega_0|^2/(\gamma\Gamma)$ (above the upper critical value $I_\mathrm{c} = |\Omega_\mathrm{0c}|^2/(\gamma\Gamma) = 531.4125$). Parameters are the same as in Fig.~\ref{Fig4}(b). The filled circles are data points, while the solid line represents a power-law fit given by $\tau\gamma = 6.571\times 10^3 (I_\mathrm{0} - I_\mathrm{c})^{-0.505}$.
(b) -- Log-log plots of $\tau$-vs-$I_0$ dependences calculated for different center-to-center distances $d$ between the SQD and MNP and for different bare detunings $\Delta$. The other parameters are the same as as in Fig.~\ref{Fig4}(b). The symbols are the data points, while the lines represent the best fits to straight lines, given by $\tau\gamma = 8.163\times 10^{3} (I_\mathrm{0} - I_\mathrm{c})^{-0.499}$ (blue line), $\tau\gamma = 6.571\times 10^3 (I_\mathrm{0} - I_\mathrm{c})^{-0.505}$ (red line), $\tau\gamma=5.063\times10^{2} (I_\mathrm{0} - I_\mathrm{c})^{-0.521}$ (black line), and $\tau\gamma=3.338\times10^{2}(I_\mathrm{0} - I_\mathrm{c})^{-0.511}$ (green line).}
\label{Fig6}
\end{figure}

We now turn to the switching time $\tau$ between the stable branches of the bistable $Z$-vs-$I_\mathrm{0}$ characteristics in the vicinity of the switching points [starting point of red and blue dashed lines in Fig.~\ref{Fig4}(b), respectively].  It is not only of fundamental interest to investigate the switching dynamics in this nonlinear system, it is also of importance in order to assess the potential usefulness of such systems as building blocks of real devices. Figs.~\ref{Fig5} and~\ref{Fig6} show the results obtained for the upper critical point. We defined $\tau$ as the time which it takes for the population difference $Z$ to acquire its first maximum after suddenly switching the incident intensity, $I_\mathrm{0} = |\Omega_0|^2/(\gamma\Gamma)$, from zero to a value slightly larger than the critical one, $I_\mathrm{c} = |\Omega_\mathrm{0c}|^2/(\gamma\Gamma)$. From Fig.~\ref{Fig5}, it is clearly seen that $\tau$ sensitively  depends on the excess of $I_\mathrm{0}$ over $I_\mathrm{c} = 531.4125$ [the latter is calculated for the set of parameters as used in Fig.~\ref{Fig4}(b)]: the system response drastically slows down when the driving intensity $I_\mathrm{0}$ approaches the critical value $I_\mathrm{c}$. Without showing details, we note that if the incident intensity is below the upper critical point, the population difference $Z$ relaxes from its initial value $Z = -1$ to the lower stable branch approximately in an exponential fashion with a time roughly on the order of the population relaxation time $\gamma^{-1}$. The switching down, from the upper stable branch to the lower one, demonstrate almost the same behavior. We do not present any calculations of these two regimes.   

In Fig.~\ref{Fig6}(a), we plotted the dependence of $\tau$ (defined as explained above) on the excess $I_\mathrm{0} - I_\mathrm{c}$ of the incident intensity $I_\mathrm{0}$ over the critical one $I_\mathrm{c}$. The numerical data points (symbols) can be well fitted by the formula $\tau\gamma = 6.571\times 10^3 (I_\mathrm{0} - I_\mathrm{c})^{-0.505}$. It is of interest to establish whether the exponent in the $\tau$-vs-$I_\mathrm{0}$ dependence, approximately equal to 0.5, is universal. In order to investigate this, we performed a series of calculations of the intensity dependence of the population relaxation time $\tau$ close to the high-intensity switching point, varying the inter-particle center-to-center distance $d$ (the coupling parameter $G$, in other words) and the off-resonance detuning $\Delta$. The results (in log-log scale) are presented in Fig.~\ref{Fig6}(b). As is seen from the numerical data and fits, the exponent $\approx 0.5$ indeed seems to be universal. The slight deviation of the data points from a straight line may be a consequence of the definition of the relaxation time $\tau$ as the time the population difference $Z$ acquires its first maximum after switching on the incident field (see Fig.~\ref{Fig5}).

The kinetics of the population difference $Z$ in the vicinity  to the lower critical point of the $Z$-vs-$I_\mathrm{0}$ bistable characteristics also show an oscillatory behavior, but not as sharp as in the vicinity of the upper critical point. Therefore, the definition of the switching time $\tau$ used above is not useful in this case. This can be understood from the fact that here the driving intensity is too low to support a population difference close to zero. Thus, the population relaxation time is dominated by the radiative decay.

\section{Summary}
\label{summary}

We conducted a theoretical study of the optical response of a heterodimer comprised of a closely spaced spherical semiconductor quantum dot and a metal nanosphere coupled to each other by dipole-dipole forces. The coupling results in a self-action of the SQD via the MNP, characterized by a complex coupling constant $G = G_\mathrm{R} + iG_\mathrm{I}$, which causes the SQD transition frequency (through $G_\mathrm{R}$) and dephasing rate (through $G_\mathrm{I}$) to depend on the SQD excited state population. This provides a feedback mechanism resulting in  bistable optical response of the system (an S-shaped behavior of the SQD population difference $Z$ versus incident intensity $I_0$).

The different physical meanings of the coupling constants $G_\mathrm{R}$ and $G_\mathrm{I}$ imply two different mechanisms of the SQD bistability. If $G_\mathrm{I} = 0$, the feedback is provided by the population-dependent resonance frequency of the SQD, while at $G_\mathrm{R} = 0$, it originates from the destructive interference of the incoming field with the secondary field produced by the SQD. Therefore, the thresholds for bistability to occur are different in these two cases: at $G_\mathrm{I} = 0$, the threshold is $G_\mathrm{R} = 4\Gamma$, whereas at $G_\mathrm{R} = 0$, it is $G_\mathrm{I} = 8\Gamma$. When both constants, $G_\mathrm{R}$ and $G_\mathrm{I}$, are not zero, the two mechanisms of bistability interfere, resulting in a quite complicated behavior of the bistability threshold as a function  of $G_\mathrm{R}$ and $G_\mathrm{I}$. We calculated the bistability phase diagrams within the system's parameter space: $G_\mathrm{R}$, $G_\mathrm{I}$ and $\Delta$ to uncover this behavior. Computations performed for a heterodimer comprised of a CdSe quantum dot and an Au nanoparticle show that bistable behavior may occur for realistic nano-particle sizes.

It should be noticed that a dimer comprised of strongly coupled two-level molecules can not manifest bistability because of the discreteness of the system's levels. The latter, in particular, prevents a continuous change of the transition frequency, which is a necessary ingredient to achieve positive feedback.~\cite{malyshev1998PRA} Thus, a SQD-MNP heterodimer is a unique nanoscopic system exhibiting this feature. Rayleigh scattering can be used as a tool to measure the effect:\cite{malyshev2011PRB} its intensity is proportional to the modulus squared of the heterodimer dipole moment. 

Having performed the steady state analysis of the optical response of a SQD-MNP heterodimer, we also studied the time it takes for the system to switch from one stable state to the other as a function of the excess of the incident intensity $I_0$ with respect to the critical (switching) value $I_c$. At the upper critical intensity, we found a power-law dependence, which surprisingly, has a universal exponent. The switching time diverges when $I_0$ approaches $I_c$, indicating the critical slowing down of the system response. The switching time in the vicinity of the lower critical point does not show such pecularities and is on the order of the population relaxation time $\gamma^{-1}$. 
 
The model we considered is the simplest hybrid nanodimer. We expect, however, that more complicated clusters (such as a SQD surrounded by several MNPs, as reported in Ref.~\onlinecite{govorov2006NL}) or a single quantum dot (or quantum dot lattice)  on top of a metal surface, can also exhibit the effects mentioned above; the MNPs just play the role of ``resonant mirrors'' and provide the positive feedback which is one of the essential ingredients for bistability to occur. 

\begin{acknowledgments}
\label{ack}

This work has been supported by NanoNextNL, a micro- and nano-technology consortium of the Government of the Netherlands and 130 partners.
\end{acknowledgments}


\begin{thebibliography}{45}
\expandafter\ifx\csname natexlab\endcsname\relax\def\natexlab#1{#1}\fi
\expandafter\ifx\csname bibnamefont\endcsname\relax
  \def\bibnamefont#1{#1}\fi
\expandafter\ifx\csname bibfnamefont\endcsname\relax
  \def\bibfnamefont#1{#1}\fi
\expandafter\ifx\csname citenamefont\endcsname\relax
  \def\citenamefont#1{#1}\fi
\expandafter\ifx\csname url\endcsname\relax
  \def\url#1{\texttt{#1}}\fi
\expandafter\ifx\csname urlprefix\endcsname\relax\def\urlprefix{URL }\fi
\providecommand{\bibinfo}[2]{#2}
\providecommand{\eprint}[2][]{\url{#2}}

\bibitem[{\citenamefont{McCall}(1974)}]{mccall1974PRA}
\bibinfo{author}{\bibfnamefont{S.~L.} \bibnamefont{McCall}},
  \bibinfo{journal}{Phys. Rev. A} \textbf{\bibinfo{volume}{9}},
  \bibinfo{pages}{1515} (\bibinfo{year}{1974}).

\bibitem[{\citenamefont{Gibbs et~al.}(1976)\citenamefont{Gibbs, McCall, and
  Venkatesan}}]{gibbs1976PRL}
\bibinfo{author}{\bibfnamefont{H.}~\bibnamefont{Gibbs}},
  \bibinfo{author}{\bibfnamefont{S.}~\bibnamefont{McCall}}, \bibnamefont{and}
  \bibinfo{author}{\bibfnamefont{T.}~\bibnamefont{Venkatesan}},
  \bibinfo{journal}{Phys. Rev. Lett.} \textbf{\bibinfo{volume}{36}},
  \bibinfo{pages}{1135} (\bibinfo{year}{1976}).

\bibitem[{\citenamefont{Lugiato}(1984)}]{lugiantoLA1984PO}
\bibinfo{author}{\bibfnamefont{L.~A.} \bibnamefont{Lugiato}},
  \bibinfo{journal}{Progress in optics} \textbf{\bibinfo{volume}{21}},
  \bibinfo{pages}{69} (\bibinfo{year}{1984}).

\bibitem[{\citenamefont{Gibbs}(1985)}]{gibbs1985B}
\bibinfo{author}{\bibfnamefont{H.~M.} \bibnamefont{Gibbs}},
  \emph{\bibinfo{title}{Optical Bistability: Controlling Light With Light}}
  (\bibinfo{publisher}{Academic Press, Inc., Orlando, FL},
  \bibinfo{year}{1985}).

\bibitem[{\citenamefont{Rosanov}(1996)}]{rosanov1996PO}
\bibinfo{author}{\bibfnamefont{N.~N.} \bibnamefont{Rosanov}},
  \bibinfo{journal}{Progress in Optics} \textbf{\bibinfo{volume}{35}},
  \bibinfo{pages}{1} (\bibinfo{year}{1996}).

\bibitem[{\citenamefont{Gibbs et~al.}(1979)\citenamefont{Gibbs, McCall,
  Venkatesan, Gossard, Passner, and Wiegmann}}]{gibbs1979APL}
\bibinfo{author}{\bibfnamefont{H.~M.} \bibnamefont{Gibbs}},
  \bibinfo{author}{\bibfnamefont{S.~L.} \bibnamefont{McCall}},
  \bibinfo{author}{\bibfnamefont{T.~N.~C.} \bibnamefont{Venkatesan}},
  \bibinfo{author}{\bibfnamefont{A.~C.} \bibnamefont{Gossard}},
  \bibinfo{author}{\bibfnamefont{A.}~\bibnamefont{Passner}}, \bibnamefont{and}
  \bibinfo{author}{\bibfnamefont{W.}~\bibnamefont{Wiegmann}},
  \bibinfo{journal}{Appl. Phys. Lett.} \textbf{\bibinfo{volume}{35}},
  \bibinfo{pages}{451} (\bibinfo{year}{1979}).

\bibitem[{\citenamefont{Kawaguchi et~al.}(1987)\citenamefont{Kawaguchi, Tani,
  and Inoue}}]{kawaguchi1987OL}
\bibinfo{author}{\bibfnamefont{H.}~\bibnamefont{Kawaguchi}},
  \bibinfo{author}{\bibfnamefont{H.}~\bibnamefont{Tani}}, \bibnamefont{and}
  \bibinfo{author}{\bibfnamefont{K.}~\bibnamefont{Inoue}},
  \bibinfo{journal}{Opt. Lett.} \textbf{\bibinfo{volume}{12}},
  \bibinfo{pages}{513} (\bibinfo{year}{1987}).

\bibitem[{\citenamefont{Gurioli et~al.}(2004)\citenamefont{Gurioli, Cavigli,
  Khitrova, and Gibbs}}]{gurioli2004PSS}
\bibinfo{author}{\bibfnamefont{M.}~\bibnamefont{Gurioli}},
  \bibinfo{author}{\bibfnamefont{L.}~\bibnamefont{Cavigli}},
  \bibinfo{author}{\bibfnamefont{G.}~\bibnamefont{Khitrova}}, \bibnamefont{and}
  \bibinfo{author}{\bibfnamefont{H.}~\bibnamefont{Gibbs}},
  \bibinfo{journal}{Phys. Stat. Sol. (a)} \textbf{\bibinfo{volume}{201}},
  \bibinfo{pages}{661} (\bibinfo{year}{2004}).

\bibitem[{\citenamefont{Cavigli and Gurioli}(2005)}]{cavigli2005PRB}
\bibinfo{author}{\bibfnamefont{L.}~\bibnamefont{Cavigli}} \bibnamefont{and}
  \bibinfo{author}{\bibfnamefont{M.}~\bibnamefont{Gurioli}},
  \bibinfo{journal}{Phys. Rev. B} \textbf{\bibinfo{volume}{71}},
  \bibinfo{pages}{035317} (\bibinfo{year}{2005}).

\bibitem[{\citenamefont{Klugkist et~al.}(2007)\citenamefont{Klugkist, Malyshev,
  and Knoester}}]{klugkist2007JCP}
\bibinfo{author}{\bibfnamefont{J.~A.} \bibnamefont{Klugkist}},
  \bibinfo{author}{\bibfnamefont{V.~A.} \bibnamefont{Malyshev}},
  \bibnamefont{and} \bibinfo{author}{\bibfnamefont{J.}~\bibnamefont{Knoester}},
  \bibinfo{journal}{J. Chem. Phys.} \textbf{\bibinfo{volume}{127}},
  \bibinfo{pages}{164705} (\bibinfo{year}{2007}).

\bibitem[{\citenamefont{Soljacic et~al.}(2003)\citenamefont{Soljacic, Ibanescu,
  Luo, Johnson, Fan, Fink, and Joannopoulos}}]{soljacic2003PRO}
\bibinfo{author}{\bibfnamefont{M.}~\bibnamefont{Soljacic}},
  \bibinfo{author}{\bibfnamefont{M.}~\bibnamefont{Ibanescu}},
  \bibinfo{author}{\bibfnamefont{C.}~\bibnamefont{Luo}},
  \bibinfo{author}{\bibfnamefont{S.~G.} \bibnamefont{Johnson}},
  \bibinfo{author}{\bibfnamefont{S.}~\bibnamefont{Fan}},
  \bibinfo{author}{\bibfnamefont{Y.}~\bibnamefont{Fink}}, \bibnamefont{and}
  \bibinfo{author}{\bibfnamefont{J.~D.} \bibnamefont{Joannopoulos}}, in
  \emph{\bibinfo{booktitle}{Proceedings of SPIE}} (\bibinfo{year}{2003}), vol.
  \bibinfo{volume}{5000}, p. \bibinfo{pages}{200}.

\bibitem[{\citenamefont{Wurtz et~al.}(2006)\citenamefont{Wurtz, Pollard, and
  Zayats}}]{wurtz2006PRL}
\bibinfo{author}{\bibfnamefont{G.~A.} \bibnamefont{Wurtz}},
  \bibinfo{author}{\bibfnamefont{R.}~\bibnamefont{Pollard}}, \bibnamefont{and}
  \bibinfo{author}{\bibfnamefont{A.~V.} \bibnamefont{Zayats}},
  \bibinfo{journal}{Phys. Rev. Lett.} \textbf{\bibinfo{volume}{97}},
  \bibinfo{pages}{57402} (\bibinfo{year}{2006}).

\bibitem[{\citenamefont{Litchinitser et~al.}(2007)\citenamefont{Litchinitser,
  Gabitov, Maimistov, and Shalaev}}]{litchinitser2007OL}
\bibinfo{author}{\bibfnamefont{N.~M.} \bibnamefont{Litchinitser}},
  \bibinfo{author}{\bibfnamefont{I.~R.} \bibnamefont{Gabitov}},
  \bibinfo{author}{\bibfnamefont{A.~I.} \bibnamefont{Maimistov}},
  \bibnamefont{and} \bibinfo{author}{\bibfnamefont{V.~M.}
  \bibnamefont{Shalaev}}, \bibinfo{journal}{Opt. Lett.}
  \textbf{\bibinfo{volume}{32}}, \bibinfo{pages}{151} (\bibinfo{year}{2007}).

\bibitem[{\citenamefont{Artuso and Bryant}(2008)}]{artuso2008NL}
\bibinfo{author}{\bibfnamefont{R.~D.} \bibnamefont{Artuso}} \bibnamefont{and}
  \bibinfo{author}{\bibfnamefont{G.~W.} \bibnamefont{Bryant}},
  \bibinfo{journal}{Nano Lett.} \textbf{\bibinfo{volume}{8}},
  \bibinfo{pages}{2106} (\bibinfo{year}{2008}).

\bibitem[{\citenamefont{Artuso and Bryant}(2010)}]{artuso2010PRB}
\bibinfo{author}{\bibfnamefont{R.~D.} \bibnamefont{Artuso}} \bibnamefont{and}
  \bibinfo{author}{\bibfnamefont{G.~W.} \bibnamefont{Bryant}},
  \bibinfo{journal}{Phys. Rev. B} \textbf{\bibinfo{volume}{82}},
  \bibinfo{pages}{195419} (\bibinfo{year}{2010}).

\bibitem[{\citenamefont{Malyshev and Malyshev}(2011)}]{malyshev2011PRB}
\bibinfo{author}{\bibfnamefont{A.~V.} \bibnamefont{Malyshev}} \bibnamefont{and}
  \bibinfo{author}{\bibfnamefont{V.~A.} \bibnamefont{Malyshev}},
  \bibinfo{journal}{Phys. Rev. B} \textbf{\bibinfo{volume}{84}},
  \bibinfo{pages}{035314} (\bibinfo{year}{2011}).

\bibitem[{\citenamefont{Brolo et~al.}(2006)\citenamefont{Brolo, Kwok, Cooper,
  Moffitt, Wang, Gordon, Riordon, and Kavanagh}}]{brolo2006PCB}
\bibinfo{author}{\bibfnamefont{A.~G.} \bibnamefont{Brolo}},
  \bibinfo{author}{\bibfnamefont{S.~C.} \bibnamefont{Kwok}},
  \bibinfo{author}{\bibfnamefont{M.~D.} \bibnamefont{Cooper}},
  \bibinfo{author}{\bibfnamefont{M.~G.} \bibnamefont{Moffitt}},
  \bibinfo{author}{\bibfnamefont{C.~W.} \bibnamefont{Wang}},
  \bibinfo{author}{\bibfnamefont{R.}~\bibnamefont{Gordon}},
  \bibinfo{author}{\bibfnamefont{J.}~\bibnamefont{Riordon}}, \bibnamefont{and}
  \bibinfo{author}{\bibfnamefont{K.~L.} \bibnamefont{Kavanagh}},
  \bibinfo{journal}{J. Phys. Chem. B} \textbf{\bibinfo{volume}{110}},
  \bibinfo{pages}{8307} (\bibinfo{year}{2006}).

\bibitem[{\citenamefont{Viste et~al.}(2010)\citenamefont{Viste, Plain, Jaffiol,
  Vial, Adam, and Royer}}]{viste2010ACN}
\bibinfo{author}{\bibfnamefont{P.}~\bibnamefont{Viste}},
  \bibinfo{author}{\bibfnamefont{J.}~\bibnamefont{Plain}},
  \bibinfo{author}{\bibfnamefont{R.}~\bibnamefont{Jaffiol}},
  \bibinfo{author}{\bibfnamefont{A.}~\bibnamefont{Vial}},
  \bibinfo{author}{\bibfnamefont{P.~M.} \bibnamefont{Adam}}, \bibnamefont{and}
  \bibinfo{author}{\bibfnamefont{P.}~\bibnamefont{Royer}},
  \bibinfo{journal}{ACS nano} \textbf{\bibinfo{volume}{4}},
  \bibinfo{pages}{759} (\bibinfo{year}{2010}).

\bibitem[{\citenamefont{Neogi and Morko{\c{c}}}(2004)}]{neogi2004NT}
\bibinfo{author}{\bibfnamefont{A.}~\bibnamefont{Neogi}} \bibnamefont{and}
  \bibinfo{author}{\bibfnamefont{H.}~\bibnamefont{Morko{\c{c}}}},
  \bibinfo{journal}{Nanotechnology} \textbf{\bibinfo{volume}{15}},
  \bibinfo{pages}{1252} (\bibinfo{year}{2004}).

\bibitem[{\citenamefont{Govorov et~al.}(2006)\citenamefont{Govorov, Bryant,
  Zhang, Skeini, Lee, Kotov, Slocik, and Naik}}]{govorov2006NL}
\bibinfo{author}{\bibfnamefont{A.~O.} \bibnamefont{Govorov}},
  \bibinfo{author}{\bibfnamefont{G.~W.} \bibnamefont{Bryant}},
  \bibinfo{author}{\bibfnamefont{W.}~\bibnamefont{Zhang}},
  \bibinfo{author}{\bibfnamefont{T.}~\bibnamefont{Skeini}},
  \bibinfo{author}{\bibfnamefont{J.}~\bibnamefont{Lee}},
  \bibinfo{author}{\bibfnamefont{N.~A.} \bibnamefont{Kotov}},
  \bibinfo{author}{\bibfnamefont{J.~M.} \bibnamefont{Slocik}},
  \bibnamefont{and} \bibinfo{author}{\bibfnamefont{R.~R.} \bibnamefont{Naik}},
  \bibinfo{journal}{Nano Lett.} \textbf{\bibinfo{volume}{6}},
  \bibinfo{pages}{984} (\bibinfo{year}{2006}).

\bibitem[{\citenamefont{Neogi et~al.}(2005)\citenamefont{Neogi, Morko{\c{c}},
  Kuroda, and Tackeuchi}}]{neogi2005OL}
\bibinfo{author}{\bibfnamefont{A.}~\bibnamefont{Neogi}},
  \bibinfo{author}{\bibfnamefont{H.}~\bibnamefont{Morko{\c{c}}}},
  \bibinfo{author}{\bibfnamefont{T.}~\bibnamefont{Kuroda}}, \bibnamefont{and}
  \bibinfo{author}{\bibfnamefont{A.}~\bibnamefont{Tackeuchi}},
  \bibinfo{journal}{Opt. Lett.} \textbf{\bibinfo{volume}{30}},
  \bibinfo{pages}{93} (\bibinfo{year}{2005}).

\bibitem[{\citenamefont{Pons et~al.}(2007)\citenamefont{Pons, Medintz,
  Sapsford, Higashiya, Grimes, Doug, and Mattoussi}}]{pons2007NL}
\bibinfo{author}{\bibfnamefont{T.}~\bibnamefont{Pons}},
  \bibinfo{author}{\bibfnamefont{I.~L.} \bibnamefont{Medintz}},
  \bibinfo{author}{\bibfnamefont{K.~E.} \bibnamefont{Sapsford}},
  \bibinfo{author}{\bibfnamefont{S.}~\bibnamefont{Higashiya}},
  \bibinfo{author}{\bibfnamefont{A.~F.} \bibnamefont{Grimes}},
  \bibinfo{author}{\bibfnamefont{S.}~\bibnamefont{Doug}}, \bibnamefont{and}
  \bibinfo{author}{\bibfnamefont{H.}~\bibnamefont{Mattoussi}},
  \bibinfo{journal}{Nano Lett.} \textbf{\bibinfo{volume}{7}},
  \bibinfo{pages}{3157} (\bibinfo{year}{2007}).

\bibitem[{\citenamefont{Zhang et~al.}(2006)\citenamefont{Zhang, Govorov, and
  Bryant}}]{zhang2006PRL}
\bibinfo{author}{\bibfnamefont{W.}~\bibnamefont{Zhang}},
  \bibinfo{author}{\bibfnamefont{A.~O.} \bibnamefont{Govorov}},
  \bibnamefont{and} \bibinfo{author}{\bibfnamefont{G.~W.}
  \bibnamefont{Bryant}}, \bibinfo{journal}{Phys. Rev. Lett.}
  \textbf{\bibinfo{volume}{97}}, \bibinfo{pages}{146804}
  (\bibinfo{year}{2006}).

\bibitem[{\citenamefont{Kosionis et~al.}(2012)\citenamefont{Kosionis, Terzis,
  Yannopapas, and Paspalakis}}]{kosionis2012JPC}
\bibinfo{author}{\bibfnamefont{S.}~\bibnamefont{Kosionis}},
  \bibinfo{author}{\bibfnamefont{A.}~\bibnamefont{Terzis}},
  \bibinfo{author}{\bibfnamefont{V.}~\bibnamefont{Yannopapas}},
  \bibnamefont{and}
  \bibinfo{author}{\bibfnamefont{E.}~\bibnamefont{Paspalakis}},
  \bibinfo{journal}{J. Phys. Chem. C} \textbf{\bibinfo{volume}{116}},
  \bibinfo{pages}{23663} (\bibinfo{year}{2012}).

\bibitem[{\citenamefont{Sadeghi}(2010{\natexlab{a}})}]{sadeghi2010NTa}
\bibinfo{author}{\bibfnamefont{S.~M.} \bibnamefont{Sadeghi}},
  \bibinfo{journal}{Nanotechnology} \textbf{\bibinfo{volume}{21}},
  \bibinfo{pages}{455401} (\bibinfo{year}{2010}{\natexlab{a}}).

\bibitem[{\citenamefont{Sadeghi}(2009{\natexlab{a}})}]{sadeghi2009PRB}
\bibinfo{author}{\bibfnamefont{S.~M.} \bibnamefont{Sadeghi}},
  \bibinfo{journal}{Phys. Rev. B} \textbf{\bibinfo{volume}{79}},
  \bibinfo{pages}{233309} (\bibinfo{year}{2009}{\natexlab{a}}).

\bibitem[{\citenamefont{Sadeghi}(2009{\natexlab{b}})}]{sadeghi2009NT}
\bibinfo{author}{\bibfnamefont{S.~M.} \bibnamefont{Sadeghi}},
  \bibinfo{journal}{Nanotechnology} \textbf{\bibinfo{volume}{20}},
  \bibinfo{pages}{225401} (\bibinfo{year}{2009}{\natexlab{b}}).

\bibitem[{\citenamefont{Sadeghi}(2010{\natexlab{b}})}]{sadeghi2010NTb}
\bibinfo{author}{\bibfnamefont{S.~M.} \bibnamefont{Sadeghi}},
  \bibinfo{journal}{Nanotechnology} \textbf{\bibinfo{volume}{21}},
  \bibinfo{pages}{355501} (\bibinfo{year}{2010}{\natexlab{b}}).

\bibitem[{\citenamefont{Sadeghi}(2012)}]{sadeghi2012APL}
\bibinfo{author}{\bibfnamefont{S.}~\bibnamefont{Sadeghi}},
  \bibinfo{journal}{Appl. Phys. Lett.} \textbf{\bibinfo{volume}{101}},
  \bibinfo{pages}{213102} (\bibinfo{year}{2012}).

\bibitem[{\citenamefont{Hatef et~al.}(2012)\citenamefont{Hatef, Sadeghi, and
  Singh}}]{hatef2012Nano}
\bibinfo{author}{\bibfnamefont{A.}~\bibnamefont{Hatef}},
  \bibinfo{author}{\bibfnamefont{S.}~\bibnamefont{Sadeghi}}, \bibnamefont{and}
  \bibinfo{author}{\bibfnamefont{M.}~\bibnamefont{Singh}},
  \bibinfo{journal}{Nanotechnology} \textbf{\bibinfo{volume}{23}},
  \bibinfo{pages}{065701} (\bibinfo{year}{2012}).

\bibitem[{\citenamefont{Ant{\'o}n et~al.}(2012)\citenamefont{Ant{\'o}n,
  Carre{\~n}o, Melle, Calder{\'o}n, Cabrera-Granado, Cox, and
  Singh}}]{anton2012PRB}
\bibinfo{author}{\bibfnamefont{M.}~\bibnamefont{Ant{\'o}n}},
  \bibinfo{author}{\bibfnamefont{F.}~\bibnamefont{Carre{\~n}o}},
  \bibinfo{author}{\bibfnamefont{S.}~\bibnamefont{Melle}},
  \bibinfo{author}{\bibfnamefont{O.}~\bibnamefont{Calder{\'o}n}},
  \bibinfo{author}{\bibfnamefont{E.}~\bibnamefont{Cabrera-Granado}},
  \bibinfo{author}{\bibfnamefont{J.}~\bibnamefont{Cox}}, \bibnamefont{and}
  \bibinfo{author}{\bibfnamefont{M.}~\bibnamefont{Singh}},
  \bibinfo{journal}{Physical Review B} \textbf{\bibinfo{volume}{86}},
  \bibinfo{pages}{155305} (\bibinfo{year}{2012}).

\bibitem[{\citenamefont{Bohren and Huffman}(1983)}]{Bohren1983B}
\bibinfo{author}{\bibfnamefont{C.~F.} \bibnamefont{Bohren}} \bibnamefont{and}
  \bibinfo{author}{\bibfnamefont{D.~R.} \bibnamefont{Huffman}},
  \emph{\bibinfo{title}{Absorption and Scattering of Light by Small Particles}}
  (\bibinfo{publisher}{J Wiley \& Sons, New York}, \bibinfo{year}{1983}).

\bibitem[{\citenamefont{Maier}(2007)}]{maier2007B}
\bibinfo{author}{\bibfnamefont{S.~A.} \bibnamefont{Maier}},
  \emph{\bibinfo{title}{Plasmonics: Fundamentals and Applications}}
  (\bibinfo{publisher}{Springer Verlag}, \bibinfo{year}{2007}).

\bibitem[{\citenamefont{Meier and Wokaun}(1983)}]{meier1983OL}
\bibinfo{author}{\bibfnamefont{M.}~\bibnamefont{Meier}} \bibnamefont{and}
  \bibinfo{author}{\bibfnamefont{A.}~\bibnamefont{Wokaun}},
  \bibinfo{journal}{Opt. Lett.} \textbf{\bibinfo{volume}{8}},
  \bibinfo{pages}{581} (\bibinfo{year}{1983}).

\bibitem[{\citenamefont{Yan et~al.}(2008)\citenamefont{Yan, Zhang, Duan, Zhao,
  and Govorov}}]{yan2008PRB}
\bibinfo{author}{\bibfnamefont{J.~Y.} \bibnamefont{Yan}},
  \bibinfo{author}{\bibfnamefont{W.}~\bibnamefont{Zhang}},
  \bibinfo{author}{\bibfnamefont{S.}~\bibnamefont{Duan}},
  \bibinfo{author}{\bibfnamefont{X.~G.} \bibnamefont{Zhao}}, \bibnamefont{and}
  \bibinfo{author}{\bibfnamefont{A.~O.} \bibnamefont{Govorov}},
  \bibinfo{journal}{Phys. Rev. B} \textbf{\bibinfo{volume}{77}},
  \bibinfo{pages}{165301} (\bibinfo{year}{2008}).

\bibitem[{\citenamefont{Hopf et~al.}(1984)\citenamefont{Hopf, Bowden, and
  Louisell}}]{hopf1984PRA}
\bibinfo{author}{\bibfnamefont{F.~A.} \bibnamefont{Hopf}},
  \bibinfo{author}{\bibfnamefont{C.~M.} \bibnamefont{Bowden}},
  \bibnamefont{and} \bibinfo{author}{\bibfnamefont{W.~H.}
  \bibnamefont{Louisell}}, \bibinfo{journal}{Phys. Rev. A}
  \textbf{\bibinfo{volume}{29}}, \bibinfo{pages}{2591} (\bibinfo{year}{1984}).

\bibitem[{\citenamefont{Ben-Aryeh et~al.}(1986)\citenamefont{Ben-Aryeh, Bowden,
  and Englund}}]{ben1986PRA}
\bibinfo{author}{\bibfnamefont{Y.}~\bibnamefont{Ben-Aryeh}},
  \bibinfo{author}{\bibfnamefont{C.~M.} \bibnamefont{Bowden}},
  \bibnamefont{and} \bibinfo{author}{\bibfnamefont{J.~C.}
  \bibnamefont{Englund}}, \bibinfo{journal}{Phys. Rev. A}
  \textbf{\bibinfo{volume}{34}}, \bibinfo{pages}{3917} (\bibinfo{year}{1986}).

\bibitem[{\citenamefont{Friedberg et~al.}(1989)\citenamefont{Friedberg,
  Hartmann, and Manassah}}]{friedberg1989PRA}
\bibinfo{author}{\bibfnamefont{R.}~\bibnamefont{Friedberg}},
  \bibinfo{author}{\bibfnamefont{S.~R.} \bibnamefont{Hartmann}},
  \bibnamefont{and} \bibinfo{author}{\bibfnamefont{J.~T.}
  \bibnamefont{Manassah}}, \bibinfo{journal}{Phys. Rev. A}
  \textbf{\bibinfo{volume}{39}}, \bibinfo{pages}{3444} (\bibinfo{year}{1989}).

\bibitem[{\citenamefont{Basharov}(1988)}]{basharov1988}
\bibinfo{author}{\bibfnamefont{A.~M.} \bibnamefont{Basharov}},
  \bibinfo{journal}{Sov. Phys. JETP} \textbf{\bibinfo{volume}{67}},
  \bibinfo{pages}{1741} (\bibinfo{year}{1988}).

\bibitem[{\citenamefont{Benedict and Trifonov}(1988)}]{benedict1988PRA}
\bibinfo{author}{\bibfnamefont{M.~G.} \bibnamefont{Benedict}} \bibnamefont{and}
  \bibinfo{author}{\bibfnamefont{E.~D.} \bibnamefont{Trifonov}},
  \bibinfo{journal}{Phys. Rev. A} \textbf{\bibinfo{volume}{38}},
  \bibinfo{pages}{2854} (\bibinfo{year}{1988}).

\bibitem[{\citenamefont{Benedict et~al.}(1991)\citenamefont{Benedict, Malyshev,
  Trifonov, and Zaitsev}}]{benedict1991PRA}
\bibinfo{author}{\bibfnamefont{M.~G.} \bibnamefont{Benedict}},
  \bibinfo{author}{\bibfnamefont{V.~A.} \bibnamefont{Malyshev}},
  \bibinfo{author}{\bibfnamefont{E.~D.} \bibnamefont{Trifonov}},
  \bibnamefont{and} \bibinfo{author}{\bibfnamefont{A.~I.}
  \bibnamefont{Zaitsev}}, \bibinfo{journal}{Phys. Rev. A}
  \textbf{\bibinfo{volume}{43}}, \bibinfo{pages}{3845} (\bibinfo{year}{1991}).

\bibitem[{\citenamefont{Malyshev et~al.}(2000)\citenamefont{Malyshev, Glaeske,
  and Feller}}]{malyshev2000JCP}
\bibinfo{author}{\bibfnamefont{V.~A.} \bibnamefont{Malyshev}},
  \bibinfo{author}{\bibfnamefont{H.}~\bibnamefont{Glaeske}}, \bibnamefont{and}
  \bibinfo{author}{\bibfnamefont{K.~H.} \bibnamefont{Feller}},
  \bibinfo{journal}{J. Chem. Phys.} \textbf{\bibinfo{volume}{113}},
  \bibinfo{pages}{1170} (\bibinfo{year}{2000}).

\bibitem[{\citenamefont{Malyshev and Moreno}(1996)}]{malyshev1996PRA}
\bibinfo{author}{\bibfnamefont{V.}~\bibnamefont{Malyshev}} \bibnamefont{and}
  \bibinfo{author}{\bibfnamefont{P.}~\bibnamefont{Moreno}},
  \bibinfo{journal}{Phys. Rev. A} \textbf{\bibinfo{volume}{53}},
  \bibinfo{pages}{416} (\bibinfo{year}{1996}).

\bibitem[{\citenamefont{Malyshev et~al.}(1998)\citenamefont{Malyshev, Glaeske,
  and Feller}}]{malyshev1998PRA}
\bibinfo{author}{\bibfnamefont{V.~A.} \bibnamefont{Malyshev}},
  \bibinfo{author}{\bibfnamefont{H.}~\bibnamefont{Glaeske}}, \bibnamefont{and}
  \bibinfo{author}{\bibfnamefont{K.~H.} \bibnamefont{Feller}},
  \bibinfo{journal}{Phys. Rev. A} \textbf{\bibinfo{volume}{58}},
  \bibinfo{pages}{670} (\bibinfo{year}{1998}).

\bibitem{endnote}Similar result has been reported in a recently published paper A.V. Malyshev, Phys. Rev. A \textbf{86}, 065804 (2012).

\bibitem[{\citenamefont{Johnson and Christy}(1972)}]{johnson1972PRB}
\bibinfo{author}{\bibfnamefont{P.~B.} \bibnamefont{Johnson}} \bibnamefont{and}
  \bibinfo{author}{\bibfnamefont{R.~W.} \bibnamefont{Christy}},
  \bibinfo{journal}{Phys. Rev. B} \textbf{\bibinfo{volume}{6}},
  \bibinfo{pages}{4370} (\bibinfo{year}{1972}).

\end{thebibliography}
\end{document}